\documentclass[preprint,12pt]{elsarticle}

\usepackage{graphicx,times}             
\usepackage{amssymb}
\usepackage{mathrsfs}
\usepackage{amsmath}
\usepackage{color}

\def\comment#1{}

\def\cT{{\cal T}}

\def\vx{\boldsymbol{x}}
\def\vn{\boldsymbol{n}}
\def\vd{\boldsymbol{d}}
\def\vw{\boldsymbol{w}}
\def\vW{{\bf W}}
\def\vv{\boldsymbol{v}}
\def\vr{\boldsymbol{r}}
\def\vg{\boldsymbol{g}}
\def\rA{{\rm A}}
\def\vX{\boldsymbol{X}}
\def\vI{\boldsymbol{I}}
\def\vz{\boldsymbol{z}}
\newcommand{\m}[1]{\mathrm{#1}}
\newcommand{\V}[1]{\boldsymbol{\m{#1}}}
\def\vbeta{\V{\beta}}

\journal{Physics Letters A}

\begin{document}

\begin{frontmatter}


\title{The gravitational time delay in the field of a slowly moving body with arbitrary multipoles }

 \author[label1,label2]{Michael H Soffel}
\ead{michael.soffel@tu-dresden.de}
 \author[label1]{Wen-Biao Han}
 \ead{wbhan@shao.ac.cn}
\address[label1]{Shanghai Astronomical Observatory, 80 Nandan Road, Shanghai, 200030, China}
\address[label2]{Lohmann Observatory, Helmholtzstrasse 10, D-01062 Dresden, Germany}

\begin{abstract}
We calculate the time delay of light in the gravitational field of a slowly moving body with arbitrary multipoles (mass and spin multipole moments) by the Time-Transfer-Function (TTF) formalism. The parameters we use, first introduced by Kopeikin for a gravitational source at rest, make the integration of the TTF very elegant and simple. Results completely coincide with expressions from the literature. The results for a moving body (with constant velocity) with complete multipole-structure are new, according to our knowledge.
\end{abstract}

\begin{keyword}
relativity \sep light propagation \sep time transfer function



\end{keyword}

\end{frontmatter}


\section{Introduction}           
\label{sect:intro}
Light propagation in gravitational fields is a very important topic not only for modern astrometry because of the high accuracies achieved in modern observations, but also
for other kinds of measurements such as radar ranging to spacecrafts or planets.
Gravitational fields cause a  propagation-time delay and a  deflection of light-rays as well as a frequency shift of the involved photons. The first effect, called Shapiro delay \cite{shapirio64}, has to be considered in space-techniques such as Very-Long-Baseline Interferometry (VLBI), Lunar Laser Ranging(LLR), and etc. The present VLBI model recommended by the IERS conventions 2010 \cite{iers2010} (the consensus model \cite{Eubanks91}) has an accuracy at the  $1$ picosecond level; it will be improved to 0.1 picosecond accuracy in the near future; LLR is approaching now the milli-meter level \cite{murphy09}. The gravitational field of the Sun produces a maximum of about 100 nanoseconds for the Earth bounded VLBI observations \cite{klioner91} and 50 nanoseconds (15 meters) in LLR experiments.

The light propagation delay in the gravitational field of a stationary mass-monopole  is quite easy to derive. For a body with arbitrary mass and spin multipole moments, moving with some velocity in the underlying coordinate system,  the treatment becomes non-trivial. People usually use the null geodesic equation to get the light-propagation information between two events (for example, emission and reception). The solutions for a gravitating body with arbitrary multipoles obtained in this way was first derived in \cite{kk92, kopeikin97}. Bertone et al. showed that the so called time-transfer-function (TTF) formalism can also be used efficiently to get the gravitational time-delay, but they dealt with the case of mass-monopoles
only \cite{ttf2014}. Recently, some authors discussed the light propagation in the field of a moving axisymmetic body \cite{Hees14}.

In this letter, we derive the TTF by means of special parameters and techniques that were first introduced by Kopeikin \cite{kk92, kopeikin97}; using this approach simplifies the calculations drastically. Results for the Shapiro-effect for a body with arbitrary mass- and spin-mutipoles are obtained in a few lines. Our results completely coincide with the ones from the literature (e.g., \cite{kk92, kopeikin97}). This calculation is then generalized in a very simple way to the case of a body moving with slow and constant velocity in the underlying coordinate system.

 In the next section, the metric of a body with arbitrary multipole-moments is recalled; in section 3 and 4, we introduce the TTF, and calculate the light propagation for the cases of arbitrary multiple moments and constant velocity. The last section contains conclusions and discussions.




\section{The Time Transfer Function}
We will consider the propagation of light-signals in a first order post-Newtonian metric of form
\begin{align}
\begin{split}\label{metform}
g_{00} &= -1+\frac{2w}{c^2} ,\\
g_{0i} &= -\frac{4}{c^3}w_i, \\
g_{ij} &= \delta_{ij}\left(1+\frac{2w}{c^2}\right) \,,
\end{split}
\end{align}
where $w$ and $w^i$ are the scalar- and the vector gravitational potentials respectively.
Our interest is in the gravitational time delay that can be computed from the null condition, $ds^2 = 0$, along the light-ray. Writing $g_{\mu\nu}=\eta_{\mu\nu}+h_{\mu\nu}$ and defining the coordinates as $(ct, x, y, z)$,  we get
\begin{displaymath}
dt^2  = {1 \over c^2} d\vx^2 + \left( h_{00} +  {2 \over c}h_{0i} {dx^i \over dt} + {1 \over c^2} h_{ij} {dx^i \over dt} {dx^j \over dt} \right) \, dt^2 \, .
\end{displaymath}
Considering $|h_{\mu\nu}| \ll 1$, to first order Taylor expansion, the above equation becomes
\begin{equation}
dt  \approx  {\vert d \vx \vert \over c} + {\vert d \vx \vert \over 2c} (h_{\mu\nu} n^\mu n^\nu) \,,
\end{equation}
where we have inserted $ {dx^i/dt} = c n^i$ from the unperturbed light-ray equation, $\vx(t) = \vx_0 + \vn c(t - t_0)$ and $n^\mu \equiv (1,\vn)$.
For our metric (\ref{metform}), the Time Transfer Function (TTF), $\cT(t_0,\vx_0;\vx) \equiv t - t_0$ with {$ds = \vert d\vx \vert$} reads
\begin{equation}\label{TTF}
 \cT(t_0,\vx_0; \vx) = \frac Rc + {1 \over 2c} \int_{s_0}^s (h_{\mu\nu} n^\mu n^\nu) ds =  \frac Rc + {2 \over c^3}
\int_{s_0}^{s} \left( w - \frac 2c \vw \cdot \vn \right) \, ds \,,
\end{equation}
where $R$ is the Euclid distance from $\vx_0$ (where a light signal is send at time $t_0$) to an observer at $\vx$ (the reception time is $t$). The TTF allows the computation of $t$ if $t_0, \vx_0$ and $\vx$ are
given. In one word, TTF is just propagation time of light in gravitational field. Because $t$ is coordinate time, the TTF as well as the time delay should be a coordinate-dependent quantity.

\section{A single gravitating body at rest}

We consider first a single body at rest at the origin of our coordinate system. Space-time outside of the body is assumed to be stationary.
Then the metric potentials outside the body take the form \cite{BD89}
\begin{align}
w &= G\sum_{l \geq 0} {(-1)^l \over l!} M_L \partial_L \left( \frac 1r \right), \label{spotential}\\
w_i &= -G\sum_{l \geq 1} {\frac{(-1)^l}{l!}\frac{l}{l+1}\varepsilon_{ijk}S_{kL-1}\partial_{jL-1}\left(\frac{1}{r}\right)} , \label{vpotential}
\end{align}
where $M_L$ and $S_L$ is the mass and spin multipole moment respectively. $L$ is a Cartesian multi-index, $L \equiv i_1 \dots i_l$ and each individual Cartesian
index $i_j$ runs over $1,2,3$ or $x,y,z$. Correspondingly, the multi-index $L-1$ indicates $l-1$ different Cartesian indices. And $r \equiv (x^2 + y^2 + z^2)^{1/2}$ is the Euclid distance from the center of mass to the field point.
We now use the Kopeikin-parametrization of  the unperturbed light-ray (see Kopeikin \cite{kopeikin97})
\begin{equation}
\vx_s = \vd + \vn \cdot s
\end{equation}
with $\vd \cdot \vn = 0$, i.e.\ $\vd = \vn \times (\vx \times \vn) = \vn \times (\vx_0 \times \vn)$ is the vector that points from the origin to the point of
closest approach of the unperturbed light-ray.
We then have $s = \vn \cdot \vx_s$ and $r_s\equiv |\vx_s| = \sqrt{d^2+s^2}$.
Following \cite{kopeikin97} we can now split the partial derivative with respect to  $x^i$ in the form
\begin{equation}
\partial_i = \partial^\perp_i + \partial^\parallel_i
\end{equation}
with
\begin{equation}
\partial^\perp_i \equiv {\partial \over \partial d_i} \, , \qquad
\partial^\parallel_i \equiv n^i {\partial \over \partial s} \, ,
\end{equation}
Then, from Eq. (24) in \cite{kopeikin97}:
\begin{equation}
\partial_L = \sum_{p=0}^l {l! \over p! (l-p)!} n^P \partial^\perp_{L-P} \partial_s^p \, ,
\end{equation}
where $n^P = n^{i_1} \dots n^{i_p}$ and $\partial_s^p = \partial^p/\partial s^p$. Inserting this into expression (\ref{TTF}) and decomposing $\cT$ as $\cT_{\rm M} +\cT_{\rm S}$ we get
\begin{align}\label{MMM}
\cT_{\rm M} = &{2G \over c^3} \sum_{l=0}^\infty \sum_{p=0}^l {(-1)^l \over l!} {l! \over p! (l-p)!}
M_L n^P \partial^\perp_{L-P} \bigg[\partial_s^p \ln {s + r \over s_0 + r_0} \nonumber \\ 
&\quad\quad\quad\quad\quad\quad\quad\quad\quad\quad\quad\quad {- \left(\partial_s^p \ln {s + r \over s_0 + r_0}\right)\Big|_{s=s_0}}\bigg]
\end{align}
for the time delay induced by the mass multipole moments $M_L$ and
\begin{align}\label{SSS}
\cT_{\rm S} =&  {4G \over c^4} \sum_{l = 1}^\infty \sum_{p=0}^l  {(-1)^l \over l!} {l! \over p! (l-p)!} {\frac{l}{l+1}}\epsilon_{ijk} n^i
S_{kL-1} n^P \partial^\perp_{jL-P-1} \nonumber \\ 
& \quad\quad\quad\quad\quad\quad\quad\quad\times\left[\partial_s^p\ln {s + r \over s_0 + r_0}{-\left(\partial_s^p \ln {s + r \over s_0 + r_0}\right)\Big|_{s=s_0}}\right]
\end{align}
for the time delay induced by the spin multipole moments $S_L$, since
\begin{equation}
{\int_{s_0}^{s}} {ds \over r_s} =  \ln {s + r \over s_0 + r_0} \, .
\end{equation}
These results are in agreement with the ones found by Kopeikin \cite{kopeikin97}.

\section{The TTF for a body slowly moving with constant velocity}

Let us now consider the situation where the gravitating body (called A) moves with a constant slow velocity $\vv_\rA$; we will neglect terms of order $v_\rA^2$ in the
following. Let us denote a canonical coordinate system moving with body A, $X^\alpha =(cT,X^a)$ (see e.g., \cite{dam1991}) and the corresponding metric potentials by
$W$ and $W^a$. The metric tensor in the co-moving system is of the form (\ref{metform}) with potentials $W,W^a$ given by Eq. (\ref{spotential}) and (\ref{vpotential}), but written in terms of co-moving coordinates.
E.g., the quantity $r$ in (\ref{spotential}) and (\ref{vpotential}) has to be replaced by $R \equiv \vert \vX \vert$,  and the spatial derivatives are now with respect to $X^a$. Under our conditions the transformation from
co-moving coordinates $X^\alpha$ to $x^\mu$ is a linear Lorentz-transformation of the form ($\vbeta_\rA \equiv \vv_\rA/c$):
\begin{equation}
x^\mu = z_\rA^\mu(T) + \Lambda^\mu_\alpha X^\alpha
\end{equation}
with $z_\rA^\mu \equiv (0,\vz_\rA(T))$ and $\Lambda^0_0 = 1, \Lambda^0_a = \beta_\rA^a, \Lambda^i_0 = \beta^i_\rA, \Lambda^i_a = \delta_{ia}$. where $\vz_{\rm A}$ is the
global coordinate position vector of body A.  A transformation of the co-moving
metric to the rest-system then yields (see also \cite{dam1991})
\begin{eqnarray} \label{potinx}
w &=& W + {4 \over c} \vbeta_\rA \cdot  \vW \nonumber \\
w_i &=& W v_\rA^i + W_i \, .
\end{eqnarray}
One can show (e.g, Zschocke \& Soffel \cite{zschocke}) that $R = r_\rA(t) + {\cal O}(v_\rA^2)$. Furthermore,
\begin{equation}
\partial_a = {\partial \over \partial X^a} = \Lambda^\mu_a {\partial \over \partial x^\mu} =
\delta_{ai} \partial_i + {\cal O}(v_\rA^2) \, ,
\end{equation}
so that the metric potential $W$ expressed in terms of $(t,\vx)$ (to first order in the velocity) takes the form
\begin{equation}
W(t,\vx) =
 G\sum_{l \geq 0} {(-1)^l \over l!} M_L \partial_L \left( \frac 1{r_\rA(t)} \right)
\end{equation}
where $M_L \partial_L = M_{i_1 \dots i_l} \partial_{i_1 \dots i_l}$ and every spatial derivative is with respect to $x^k$.
Similarly, for the gravito-magnetic potential $W^i$ one finds
\begin{equation}\label{Winx}
W_i(t,\vx)= -G\sum_{l \geq 1} {\frac{(-1)^l}{l!}\frac{l}{l+1}\varepsilon_{ijk}S_{kL-1}\partial_{jL-1}\left(\frac{1}{r_\rA(t)}\right)} ,
\end{equation}
and the TTF is given by expression (\ref{TTF}) with (\ref{potinx}) - (\ref{Winx}).

With $\vz_\rA(t) = \vz_\rA^\rA + \vv_\rA (t - t_\rA)$ we get along the unperturbed light-ray with $\vx(t) = \vx_0 + \vn c (t - t_0)$
\begin{equation}
\vr_\rA(t) = \vx_0 - \vz_\rA(t_0) + (\vn - \vbeta_\rA) c (t - t_0) \, ,
\end{equation}
i.e., due to first order aberration, the unit vector along the unperturbed light-ray, as seen from the moving body A, is given by
\begin{equation}
\vn_\beta \equiv {\vg_\beta \over g_\beta}
\end{equation}
with
\begin{equation}
\vg_\beta \equiv \vn - \vbeta_\rA \, .
\end{equation}
We can then write the TTF in the form
\begin{align}
\cT(t_0,\vx_0;\vx) &= \frac Rc + {2 \over c^3} {\int_{s_0}^{s}} \left[ W \cdot (1 - 2 \vbeta_\rA \cdot \vn) - \frac 2c
(\vn - 2 \vbeta_\rA) \cdot \vW \right]\, ds \nonumber \\
&= \frac Rc + {2 g_\beta \over c^3} {\int_{s'_0}^{s'}} W(s') ds' - {4 \over c^4} \vn_\beta \cdot {\int_{s'_0}^{s'}} \vW(s') \, ds' \nonumber \\
&\quad\quad\,\,+ {4 \over c^4} \vbeta_\rA
\cdot \int_{s'_0}^{s'} \vW(s') \, ds' \,,
\end{align}
where $s' = g_\beta s$.

We now
parametrize the unperturbed light-ray in the form
\begin{equation}
\vx_\sigma = \vz_\rA + \vd_\beta + \vn_\beta  \sigma \,,
\end{equation}
where $\vd_\beta = \vn_\beta \times(\vr_\rA \times \vn_\beta)$ is perpendicular to $\vn_\beta$ so that $r_\rA(t) = \sqrt{ d_\beta^2 + \sigma^2}$ and $\sigma = \vr_\rA \cdot \vn_\beta$.
Similar to the case of a body at rest we split the spatial derivative into to two parts, $\partial_i = \partial_i^\perp + \partial_i^\parallel$, with
\begin{equation}
\partial_i^\perp = {\partial \over \partial d_\beta^i} \,, \qquad \partial_i^\parallel = n_\beta^i {\partial \over \partial \sigma} \, .
\end{equation}
The TTF therefore takes the form
\begin{align} \label{Mmoving}
\cT_{\rm M} =&
{2 g_\beta G \over c^3} \sum_{l=0}^\infty \sum_{p=0}^l {(-1)^l \over l!} {l! \over p! (l-p)!}
M_L n^P_\beta \partial^\perp_{L-P} \bigg[\partial^p_{\sigma} \ln {r_\rA  + \sigma \over r_\rA^0 + \sigma^0} \nonumber \\ 
& \quad\quad\quad\quad\quad\quad\quad\quad\quad\quad\quad\quad\quad\quad - \left(\partial_\sigma^p \ln {r_\rA +\sigma \over r_\rA^0+\sigma^0}\right)\Big|_{\sigma=\sigma^0}\bigg]
\end{align}
for the gravitational time-delay due to the mass-multipole moments of the moving body and  
\begin{align}\label{Smoving}
\cT_{\rm S} = &{4 G \over c^4} \sum_{l \ge 1} \sum_{p=0}^l {(-1)^l \over l!} {l! \over p! (l-p)!} {l \over l+1} \epsilon_{ijk}
(n_\beta^i - \beta_A^i) S_{kL-1} n_\beta^P \partial^\perp_{jL-P-1}  \nonumber \\ 
& \quad\quad\quad\quad\quad\quad\quad\quad\quad\times\bigg[\partial^p_\sigma \ln {r_\rA + \sigma \over r_\rA^0 + \sigma^0}- \left(\partial_\sigma^p \ln {r_\rA +\sigma \over r_\rA^0+\sigma^0}\right)\Big|_{\sigma=\sigma^0}\bigg]
\end{align}
for the gravitational time-delay due to the moving spin multipoles, where $n_\beta^P = n_\beta^{i_1} \dotsb n_\beta^{i_p}$ and all the terms proportional to $\vbeta_\rA^2$ should be dropped. In this work the moving multipoles are time-independent. For the case of arbitrary time-dependent (but non-moving) multipoles, see \cite{kopeikin14}.

Let
\begin{equation}
\Phi(\sigma,\vd) \equiv \ln(\sigma + \sqrt{d^2 + \sigma^2}) \,,
\end{equation}
then the first derivatives appearing in (\ref{Mmoving}) and (\ref{Smoving}) (with $\vn$ and $\vd$ being replaced by $\vn_\beta$ and $\vd_\beta$) read:
\begin{eqnarray}
\partial_\sigma  \Phi &=&  \frac 1r \\
\partial^2_\sigma \Phi &=& - {\sigma \over r^3} \\
\partial^\perp_i \Phi &=& {d^i \over r(r+s)} \\
\partial^\perp\partial_s \Phi &=& - {d^i \over r^3} \\
\partial^{\perp}_{<ij>} \Phi &=& - {(s + 2r) \over (r+s)^2 r^3} d^i d^j - {n^i n^j \over r(r+s)} \, ,
\end{eqnarray}
where the last term results from the fact that \cite{kopeikin97}:
\begin{equation}
\partial^\perp_j d^i = \delta_{ij} - n^i n^j \, .
\end{equation}

\noindent
Considering e.g., the mass-monopole term we have
\begin{equation*}
\cT_{{\rm M}, l=0} = 2 {G M_\rA \over c^3} g_\beta \ln {r_\rA + \sigma \over r_\rA^0 + \sigma^0}
\end{equation*}
and since $\sigma = \vn_\beta \cdot \vr_\rA = \vg_\beta \cdot \vr_\rA/g_\beta$, we obtain
\begin{equation}
\cT_{{\rm M}, l=0} =  {2 G M_\rA\over c^3}g_\beta
\ln \left( {\vg_\beta \cdot \vr_\rA + g_\beta r_\rA \over \vg_\beta \cdot \vr_\rA^0 + g_\beta r_\rA^0} \right)
\end{equation}
in accordance with the results from the literature (e.g., \cite{klioner91}, \cite{ttf2014} ).

\bigskip\noindent
For the mass-quadrupole in uniform motion we get
\begin{equation} \label{quadrupolediff}
\cT_{{\rm M}, {l=2}} = {G \over c^3} g_\beta M_{ij} I_{ij}
\end{equation}
with
\begin{eqnarray}
I_{ij} &=& (n_\beta^i n_\beta^j \partial^2_\sigma + 2 n_\beta^i \partial_\sigma \partial_j^\perp + \partial^\perp_{ij}) \Phi \large\vert^s_0 \nonumber \\
&=& - n_\beta^i n_\beta^j \left( {\sigma \over r^3} + {1 \over r(r+\sigma)} \right) - {2 n^i d^j \over r^3} - {d^i d^j (\sigma + 2r) \over (r+\sigma)^2 r^3} \, .
\end{eqnarray}
Taking the integral expression for $\cT_{{\rm M}, l=2}$ one gets the same form  as in (\ref{quadrupolediff}) but with $I_{ij}$ being replaced by
\begin{eqnarray}
I'_{ij} &=&
3 \int_{\sigma_0}^\sigma {\frac{(d_\beta^i+n_\beta^i \sigma)(d_\beta^j+n_\beta^j \sigma)}{(d_\beta^2+\sigma^2)^{5/2}}d\sigma} \nonumber\\
&=&  \left(\frac{\sigma^3}{r^3}\frac{n_\beta^i n_\beta^j}{d_\beta^2}-\frac{2n_\beta^i d_\beta^j}{r^3}+\frac{3\sigma d_\beta^2+2\sigma^3}{r^3}\frac{d_\beta^i d_\beta^j}{d_\beta^4}\right)\biggr|_{\sigma_0}^{\sigma} \, .
\end{eqnarray}
With some re-writing, using $d_\beta^2 = r^2  - \sigma^2$, one finds that $I'_{ij} = I_{ij} + {\rm const.}$. Expression (\ref{quadrupolediff})  agrees with the one given by Klioner \cite{klioner91} when $\vv  = 0$.

\bigskip
The contribution from the spin-dipole can be written in the form
\begin{equation}
\cT_{{\rm S}, l = 1}
 = - {2 G \over c^4} \epsilon_{ijk} (n_\beta^i - \beta_\rA^i) I_j S_k
\end{equation}
with
\begin{equation}
I_j = \partial_j \int {d\sigma \over r_\rA^\sigma} = (\partial^\perp_j + n_\beta^j \partial_\sigma)  \ln {r_\rA  + \sigma \over r_\rA^0 + \sigma^0}
\end{equation}
or
\begin{equation}
\vI = {1 \over r_\rA} \left( \vn_\beta - {\vd_\beta \, \sigma \over d_\beta^2} \right) - {1 \over r_\rA^0} \left( \vn_\beta - {\vd_\beta \, \sigma_0 \over d_\beta^2} \right) \,.
\end{equation}
A result for the gravitational time delay caused by a moving spin-dipole has already been published by Kopeikin \& Mashhoon \cite{kopeikin02}. They have actually used the same expression
(\ref{TTF}) for the gravitational time delay and it has been shown in \cite{zschocke} that their metric is in agreement with the one used in this paper, so the results must agree, though Kopeikin \& Mashhoon used retarded quantities throughout.

\section{Conclusions}
In this letter, the Time-Transfer-Function as derived from the null condition of light in vacuum, is used to derive the gravitational time delay.

 The use of the Kopeikin-decomposition of spatial derivatives makes this method especially elegant for gravitational bodies with arbitrary (time independent) multipole moments. By introducing the first order aberration, we extend our results to a moving body with constant velocity and also arbitrary multipole moments. 

 This work was done in the frame of our efforts to formulate an exhaustive documentation of a relativistic VLBI model that could be adopted by international panels.

\emph{Acknowledgments}
M. Soffel was funded by Chinese Academy of Sciences visiting professorship of senior international scientists, GRANT No. 2013T2J0044; W.-B. Han was supported by the National Natural Science Foundation of China (NSFC) under No.11273045.


\begin{thebibliography}{99}
\bibitem{shapirio64} I.I. Shapiro, Phys. Rev. Lett., 13 (1964) 789
\bibitem{iers2010}  G. Petit, and B. Luzum (eds.), IERS Conventions 2010. (IERS Technical Note; 36) Frankfurt am Main: Verlag des Bundesamts f\"{u}r Kartographie und Geod\"{a}sie, 2010
\bibitem{Eubanks91} T.M. Eubanks (eds.), Proceedings of the U.S. Naval Observatory Workshop on Relativistic Models for Use in Space Geodesy, U.S. Naval Observatory, Washington, D.C., June, 1991
\bibitem{murphy09} T.W. Murphy,  Space Science Review, 148 (2009) 217-223
\bibitem {klioner91}S.A. Klioner, General relativistic model of VLBI observations, in Proc. AGU Chapman Conf. on Geodetic VLBI: Monitoring Global Change, W.E. Carter (ed.), NOAA Rechnical Report NOS 137 NGS 49, 1991, American Geophysical Union, Washington, D.C., 188
\bibitem{kk92} S. Klioner, S. Kopeikin, Astro. J, 104 (1992) 897
\bibitem{kopeikin97}  S. Kopeikin, J. Math. Phys., 38 (1997) 2587
\bibitem{ttf2014}  S. Bertone, O. Minazzoli, M. Crosta et al., Class. Quantum Grav., 31 (2014) 015021
\bibitem{Hees14} A. Hees, S. Bertone, C. Le Poncin-Lafitte, Phys. Rev. D 90 (2014) 084020
\bibitem{BD89}  L. Blanchet, T. Damour, Ann. Inst. Henri Poincar\'{e} A, 50 (1989) 377
\bibitem{dam1991} T. Damour, M. Soffel, C. Xu, Phys.Rev., D 43 (1991) 3273
\bibitem{zschocke} S. Zschocke, M. Soffel, Class. Quantum Grav. 31 (2014) 175001 
\bibitem{kopeikin14} P. Korobkov, S. Kopeikin, In "Frontiers in relativistic celestial mechanics", S. Kopeikin (ed.), Vol. 1, 195, Berlin, De Gruyter, 2014
\bibitem{kopeikin02}  S. Kopeikin, B. Mashhoon, Phys. Rev. D, 65 (2002) 064025
\end{thebibliography}
\end{document}